\pdfoutput=1
\documentclass[%
 reprint,
 amsmath,amssymb,
 aps,
 prd,
]{revtex4-2}

\usepackage{graphicx}
\usepackage{hyperref}
\usepackage[caption=false]{subfig}
\usepackage{multirow}
\usepackage{comment}
\usepackage{textcomp}
\usepackage[capitalise,noabbrev]{cleveref}
\crefname{section}{\S}{\S\S}
\Crefname{section}{\S}{\S\S}
\crefformat{section}{\S#2#1#3}
\crefname{paragraph}{\S}{\S\S}
\crefformat{paragraph}{\S#2#1#3}
\usepackage{bm}
\usepackage[dvipsnames]{xcolor}
\usepackage[range-phrase = --, range-units = single]{siunitx}

\newcommand{\smrm}[1]{_{\mathrm{#1}}}

\newcommand{\expLR}[1]{\ensuremath\exp{\left(#1\right)}}
\newcommand{\logLR}[1]{\ensuremath\log{\left(#1\right)}}

\newcommand{\atanLR}[1]{\ensuremath\arctan{\left(#1\right)}}

\usepackage[normalem]{ulem}

\begin{document}

\title{Feasibility study of beam-expanding telescopes in the interferometer arms for the Einstein Telescope}
\author{Samuel Rowlinson$^1$}
\author{Artemiy Dmitriev$^1$}
\author{Aaron W. Jones$^2$}
\author{Teng Zhang$^1$}
\author{Andreas Freise$^{1,3,4}$}

\affiliation{$^1$School of Physics and Astronomy, and Institute of Gravitational Wave Astronomy,University of Birmingham, Edgbaston, Birmingham B15 2TT, United Kingdom}
\affiliation{$^2$OzGrav, University of Western Australia, Crawley, Western Australia, Australia}
\affiliation{$^3$Department of Physics and Astronomy, VU Amsterdam, De Boelelaan 1081, 1081, HV, Amsterdam, The Netherlands}
\affiliation{$^4$Nikhef, Science Park 105, 1098, XG Amsterdam, The Netherlands}

\begin{abstract}
The optical design of the Einstein Telescope (ET) is based on a dual-recycled Michelson interferometer with Fabry-Perot cavities in the arms. ET will be constructed in a new infrastructure, allowing us to consider different technical implementations beyond the constraints of the current facilities. In this paper we investigate the feasibility of using beam-expander telescopes in the interferometer arms. We provide an example implementation that matches the optical layout as presented in the ET design update 2020. We further show that the beam-expander telescopes can be tuned to compensate for mode mismatches between the arm cavities and the rest of the interferometer.
\end{abstract}

\date{\today}

\maketitle

\section{Introduction}
\label{se:intro}

\begin{figure}[hb]
    \centering
    \includegraphics[width=\columnwidth]{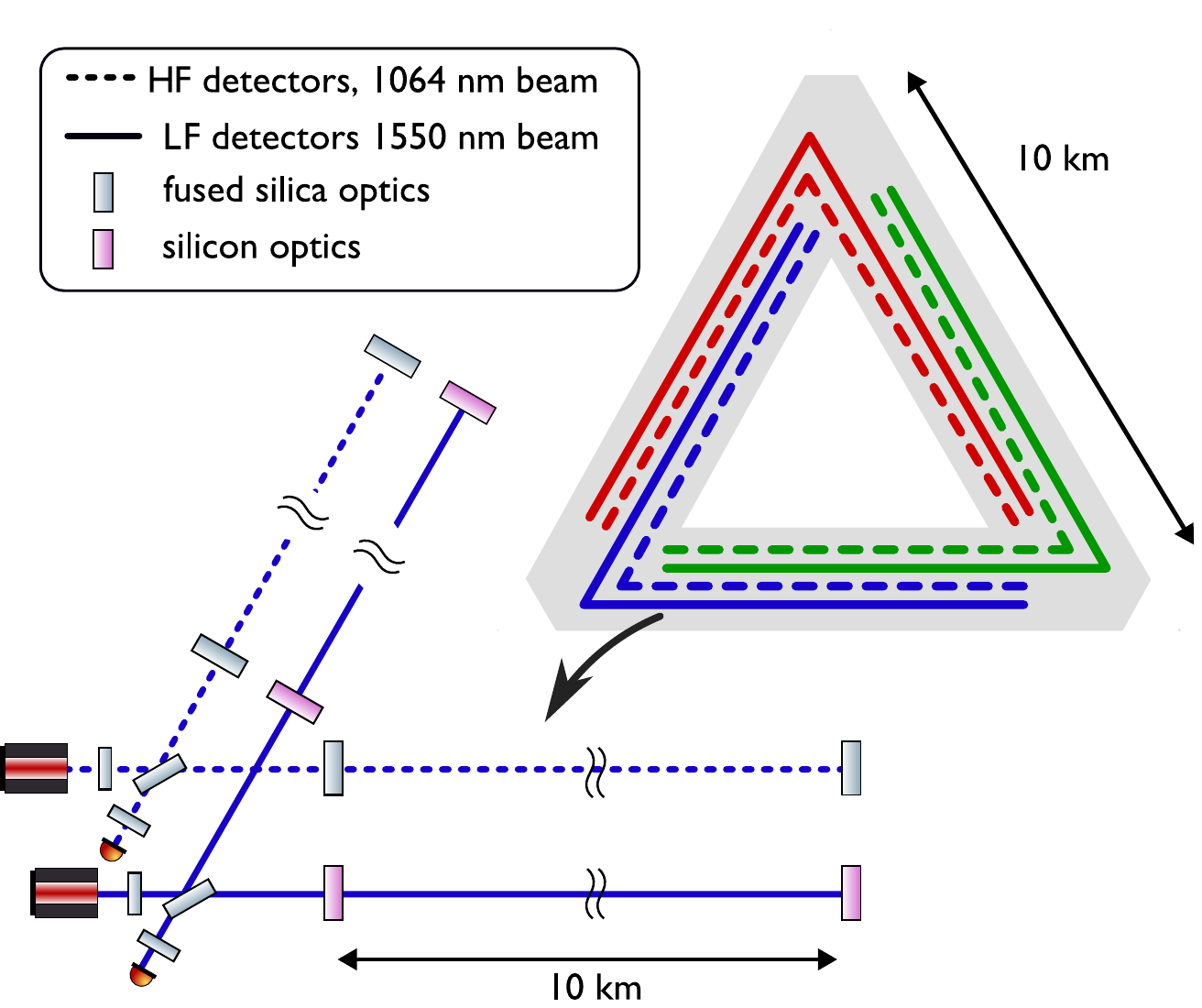}
    \caption{The diagram on the top left shows a general overview of the ET observatory layout, with 3 detectors forming  a equilateral triangle of 10\,km length. Bottom left is a sketch that shows that each detector consists of two interferometers, one optimised for high frequencies (HF) and one for low frequencies (LF). The core interferometer layout is based on a Michelson interferometer with Fabry-Perot cavities in the arms and recycling.}
    \label{fig:et_triangle}
\end{figure}

The Einstein Telescope (ET) is a proposed \emph{third-generation} gravitational wave detector~\cite{ET2020}. Once constructed, ET will provide an unprecedented level of sensitivity enabling: precise tests of General Relativity, studies of compact binary coalesces involving both intermediate black holes and neutron stars, and will be sensitive enough to test several dark matter candidates~\cite{Maggiore_2020}. ET combines a unique layout and design combining well-proven concepts from current gravitational-wave detectors with new technology. Figure~\ref{fig:et_triangle} shows a sketch with the basic features of the layout: the ET observatory is composed of three detectors that together form an equilateral triangle. Each detector consists of two interferometers, one low-frequency detector (ET-LF)
with its sensitivity optimised for low frequencies from 3\,Hz to 30\,Hz and another high-frequency detector (ET-HF) with its sensitivity optimised for high frequencies from 30\,Hz to 10\,kHz.
Similarly to current generation gravitational wave detectors, i.e. Advanced LIGO~\cite{ligo_2015} and Advanced Virgo~\cite{Acernese_2014}, each interferometer in ET is a Fabry-Perot Michelson interferometer with power and signal recycling cavities (PRC \& SRC) for arm cavity power enhancement and shaping the signal response~\cite{PhysRevD.67.062002}, respectively.  This interferometer configuration presented in~\cite{ET2020} represents the initial detector anticipated to be installed, with upgrades and refinements to be implemented over several decades. The details of the design of the initial detector will be prepared, in sync with research and development of the required technology, over the next years.

\begin{figure*}[tb]
    \centering
    \includegraphics[width=1.9\columnwidth]{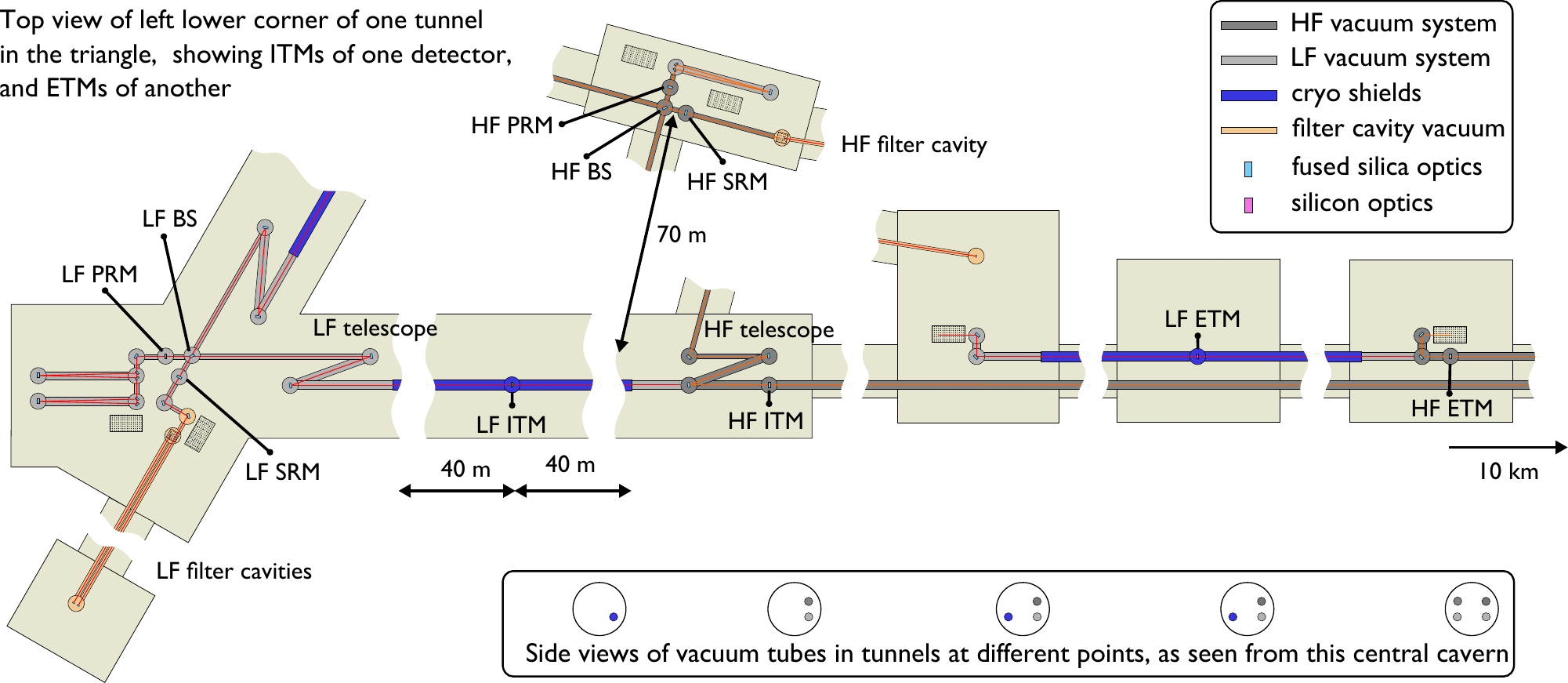}
    \caption{This is a sketch of the lower left corner of the triangle, showing an example implementation of the optical layout, the vacuum system for the main optics and the corresponding cavern layout. In particular this shows the possible location of Z-shaped telescope systems for ET-LF and ET-HF detectors. In this example the telescopes have been placed to achieve an angle of incidence of 45\, degrees on the ET-HF beamsplitter.}
    \label{fig:et_layout}
\end{figure*}

The Fabry-Perot cavities in the arms of the detectors are designed to have large beam sizes on the test mass mirrors, i.e. the input test mass (ITM) and end test mass (ETM), for reducing the impact of thermal noise of the optics on the detector sensitivity. Beam-expander telescopes are used to match input light with smaller beam diameters to the larger beams in the arm cavities. In Advanced LIGO such telescopes are located between the beamsplitter and the recycling mirrors~\cite{Arain:08}, and in Advanced Virgo similar telescopes are part of the input-output optics outside the main interferometer \cite{Buy_2017}.
However, in the ET the beam sizes on the main mirrors are significantly larger, requiring a very large substrate for the central beamsplitter, larger than the main optics, due to the angle of incidence of 60\,deg. In this paper we investigate the feasibility of an alternative layout with  beam-expander telescopes located between the main beamsplitter (BS) and the arm cavities. Such telescopes would provide
smaller beam sizes in the central interferometer formed by the beamsplitter, power recycling mirror (PRM) and signal recycling mirror (SRM), allowing for using much smaller optical components. This not only reduces the cost and complexity of these optics and their suspension systems, but also simplifies the mitigation of secondary reflections and scattered light, and reduces the effect of beam jitter~\cite{PhysRevD.95.062001}. Figure~\ref{fig:et_layout} shows a sketch of the optical layout in the lower left corner of the ET triangle, including possible locations for the beam expansion telescopes. An additional advantage of positioning the telescopes between the beamsplitter and arm cavity is that Z-shaped telescopes provide flexibility in beam steering, for example they provide the possibility to steer the ET-HF beam around the suspension system of the ET-LF ITM, and they decouple the angle of incidence on the main beamsplitter from the angle between the long interferometer arms~\cite{DeSalvo2020a, DeSalvo2020b}.

In this paper we analyse different arm telescope designs with the target of achieving a 6\,mm large beam waist at the main beamsplitter. The constraint on the size of this waist (and thus the spot size on the beamsplitter) is rationalised in \cref{se:arm_cav_mode}, where other constraints stemming from the design of the central interferometer and the arm cavities~\cite{ET2020} are discussed as well. Regions of interest in our parameter searches, covered in \cref{se:tel_designs_lf} for ET-LF and in~\cref{se:tel_designs_hf} for ET-HF, are defined as regimes where the telescope configuration gives this 6 mm waist positioned at the beamsplitter, whilst the SRC is stable.
In \cref{se:tolerancing}, we present an analysis on the sensitivity of the ET-LF telescope solution to each free parameter. Following on from this, a preliminary study quantifying the necessary active changes (for mode matching) to the telescope mirror curvatures in the presence of thermal lensing in the ITM, is given.
The results in this paper were obtained using beam parameter propagation~\cite{Kogelnik:65} via the symbolic ABCD matrix capabilities of our open-source simulation software \texttt{FINESSE}~3~\cite{finesse3}. The methodology involved grid-based searches of the parameter space to find regions satisfying the aforementioned requirements. In this context, grid-based searches are defined as analyses of high dimensional scans over any given free parameters. An independent verification of the results was performed using a new analytical framework for optimisation of beam expansion telescopes in coupled cavities~\cite{Dmitriev2020}.

\section{Arm Cavity Eigenmode and Telescope Constraints}
\label{se:arm_cav_mode}

The maximal beam sizes on test masses is set via the tolerance on power lost through clipping at the mirror edge. The relation between the maximum beam size $w\smrm{max}$ to achieve a minimum clipping loss $l_{\mathrm{clip}}$ is given by:
\begin{equation}
    w\smrm{max} = \sqrt{\frac{2}{\logLR{\frac{1}{l\smrm{clip}}}}} R\,,
    \label{eq:w_of_lclip}
\end{equation}
with $R$ as the mirror radius~\cite{IfoLivRev}.
The resulting beam sizes and corresponding mirror radii of curvature for ET-HF and ET-LF that satisfy the requirement of clipping loss of 1 ppm are listed in Table~\ref{tab:armcav_geometric_params}. Note that the diameters of the mirrors shown in this table were taken directly from~\cite{ET-D,ET2020} but do not necessarily represent final design values. The arm cavity parameters shown in Table~\ref{tab:armcav_geometric_params} are used to define the arm cavity model serving as the starting point of the beam propagation analyses in~\cref{se:tel_designs_lf}.

\begin{table}
    \caption{A summary of the key parameters of the ET-LF and ET-HF arm cavities.
    Note that we have assumed symmetric arm cavities for simplicity here.}
    \begin{center}
        \setlength{\tabcolsep}{0.5em} %
        \def\arraystretch{1.2}%
        \begin{tabular}{|c|c|c|}
            \hline
            \, & \textbf{ET-HF} & \textbf{ET-LF} \\ \hline
            Wavelength ($\lambda$) & 1064 nm & 1550 nm \\ \hline
            Cavity length ($L$) & 10 km & 10 km \\ \hline
            FSR ($\Delta\nu$) & 15 kHz & 15 kHz \\ \hline
            ITM/ETM diameter ($M_d$) & 62 cm & 45 cm \\ \hline
            ITM/ETM curvature ($R_C$) & 5070 m & 5580 m \\ \hline
            Beam radius on ITM/ETM ($w$) & 12.0 cm & 9.0 cm \\ \hline
            Beam radius at cavity waist ($w_0$) & 1.42 cm & 2.90 cm \\ \hline
            Rayleigh range ($z_R$) & 591 m & 1702 m \\ \hline
            Distance to waist from ITM ($z_0$) & 5 km & 5 km \\ \hline
            Cavity stability factor ($g$) & 0.95 & 0.63 \\ \hline
            Round-trip gouy phase ($\psi_{\mathrm{RT}}$) & 333$^{\circ}$ & 285$^{\circ}$ \\ \hline
            Mode separation frequency ($\delta f$) & 1.1 kHz & 3.1 kHz \\ \hline
        \end{tabular}
    \end{center}
    \label{tab:armcav_geometric_params}
\end{table}

\subsection{Telescope parameter constraints}
\label{se:constraints}
\begin{figure}
    \centering
    \includegraphics[width=\columnwidth]{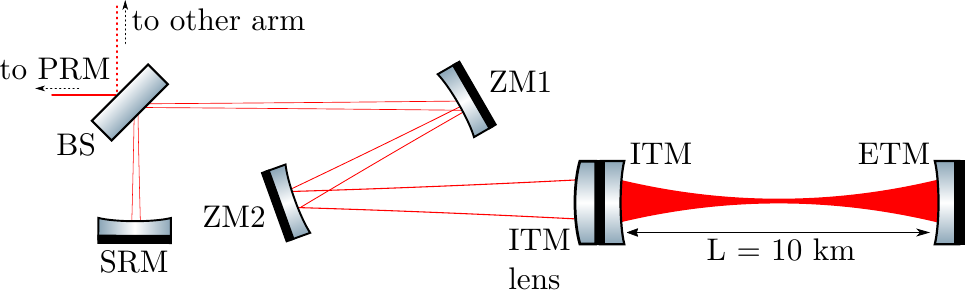}
    \caption{Schematic of an ET arm telescope with a lens at the input test mass (ITM). This type of configuration is used
    for the analyses in this section, where the Z mirrors are flat in \cref{se:flat_zms} whilst the lens has an infinite focal length in \cref{se:curved_zms_no_lens}.}
    \label{fig:telescope}
\end{figure}

A simplified schematic of one beam-expander telescope is shown in Fig.~\ref{fig:telescope}.
In order to reduce the impact of thermal noise on the sensitivity, ET-LF will make use of cryogenics to cool down the test masses to $\sim$ 10 to 20\,K.
Cryoshields of around 40\,m length will be used along the beam before and after the cryogenic mirrors~\cite{ET-D}, placing a lower limit on the distance between the telescope mirrors and the ITM. Thus, commensurately setting a lower limit for the SRC length for ET-LF. In \cref{se:curved_zms_no_lens} and \cref{se:curved_zms_with_lens} this distance is kept fixed at the current design value of 52.5\,m.

The picture is different for ET-HF, where a signal recycling cavity length of 100 m or less is preferred in order to improve the quantum-noise
limited higher-frequency sensitivity~\cite{PhysRevD.101.082002}. ET-HF does not use cryogenics, so that the lower limit on the SRC length is given only by the minimum distance allowed between the vacuum tanks.

As stated in~\cref{se:intro}, we targeted a waist size of 6\,mm at the beamsplitter. The final design for the size of the beam on the central beamsplitter will be based on a trade-off study including the following considerations. In Advanced LIGO the main beamsplitter sits between the telescopes and the arm cavities \cite{Arain:08} --- resulting in a spot size on the central beamsplitter comparable to the beam size on the ITMs; and thus requiring a large beamsplitter to avoid clipping losses. In contrast, in this work, we propose placing the telescopes between the beamsplitter and the arms, to allow use of a smaller beamsplitter, an idea briefly discussed in~\cite{Granata_2010}. Assuming a beamsplitter radius of 15\,cm and 60 degree intersection angle of two incident beams, sub part per million clipping losses are achieved with beam sizes smaller than $\sim$10\,mm, setting an upper bound on the acceptable beam size at the beamsplitter.

Similarly to GEO600 \cite{L_ck_2010}, ET-HF will operate with $\sim$\,kW levels of power on the central beamsplitter \cite{ET-D}. As such, this circulating power will induce thermal lensing in the beamsplitter substrate, as is the case for GEO600~\cite{Wittel18}. This thermal lensing causes an undesirable excitation of higher-order modes and reduces the interferometric visibility. The strength of this lensing is related to the beam intensity and thus reduced with larger beams at the central beamsplitter. For consistency between both solutions, a waist size of 6\,mm at the central beamsplitter was targeted for both ET-LF and ET-HF. It is important to note, however, that, in the case of ET-HF, this may contribute undesirably to scattering into higher-order modes. The strength of this effect will require further studies.

Note that although the targeted beam size on the beamsplitter is 15 to 20 times smaller than that on the test masses of ET-LF and ET-HF, respectively, the thermal noise contribution from the main beamsplitter is still much smaller than that from arm cavity mirrors, taking into account the arm cavity finesse, at around 900, and the fact that the beamsplitter will have fewer coating layers and a smaller substrate volume.



\section{ET-LF arm telescope design}
\label{se:tel_designs_lf}

In this section we analyse potential arm telescope configurations focusing on achieving a stable SRC whilst
keeping a waist size of $w_0 \sim 6\,\mathrm{mm}$ near to the beamsplitter. The analyses in this section are performed for ET-LF. In this and the following sections we will describe beam-expander telescopes with curved mirrors with a non-normal angle of incidence. With the commonly used spherical mirrors, such telescopes would suffer from astigmatism which would reduce the mode matching in the interferometer. Note that the following computations assume spherical mirrors with small angles of incidence and negligible astigmatism. The schematic layout shown in Figure~\ref{fig:et_layout} however implies relatively large angles of incidence; requiring non-spherical mirrors. If the final optical design will include significant angles of incidence, our results provide a good starting point for designing the required non-spherical surfaces, based on the desired beam parameters along the optical path.

\begin{figure}[b]
    \centering
    \includegraphics[width=\columnwidth]{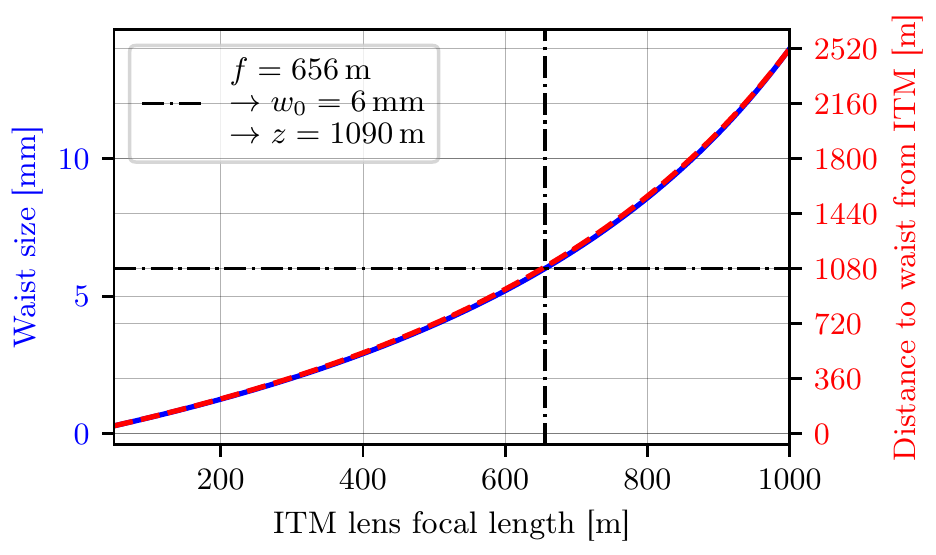}
    \caption{The waist-size (blue) and distance to waist from ITM (red) of a beam matched to the arm cavity for different ITM lens focal lengths.
    Highlighted on this plot is the focal length value which corresponds to our targeted waist size. Note that even with some flexibility on the
    waist size of a few millimeter, the required distance from the ITM to the waist is still on the order of at least 0.5\,km.}
    \label{fig:w0_vs_f_flat_zms}
\end{figure}

\subsection{Flat Z mirrors}
\label{se:flat_zms}

The simplest beam-expander configuration possible is one based on a lens at the ITM whilst keeping the
Z mirrors flat (i.e. they are just steering mirrors).
Using our target waist size, we can deduce an ITM lens focal length value for this setup, which is shown in Figure \ref{fig:w0_vs_f_flat_zms}. Also displayed in that figure are the distances to the beam waist from the ITM. This figure demonstrates that a distance of $\sim 1\,\mathrm{km}$ is necessary for maintaining a waist, of the appropriate size, at (or near to) the beamsplitter. Given the impracticality of this distance (implying
a comparable SRC length), we can reject this type of configuration.

\subsection{Curved telescope mirrors, and no lens at ITM}
\label{se:curved_zms_no_lens}

Allowing the mirrors in the Z-configuration to have some curvature gives another type of beam-expander
configuration which can be explored for feasibility. Based on the results in the single lens case
we expect ZM2 to have a positive RoC for converging the beam rapidly and ZM1 to have a negative RoC in order to achieve a small beam size on the beamsplitter over a short distance.

Another criterion comes from the stability of the SRC. The setup has significant number of free degrees of freedom, \textit{i.e.} RoCs of ZM1 and
ZM2 , the telescope distance (the distance between ZM1 and ZM2) and other free spaces in the SRC, which determine the round trip Gouy phase. But the basic behaviour of this system can be understood intuitively when combining a basic  understanding of beam propagation and our simulation results. Firstly, the accumulated Gouy phase contribution from ITM to ZM2 can be ignored, since this distance is much smaller than the Rayleigh range of the beam from the arm cavities which is $\sim$ 1.7\,km as shown in Table.~\ref{tab:armcav_geometric_params} and the Gouy phase is calculated as,
\begin{equation}\label{eq:psi}
    \psi = \atanLR{\frac{z}{z_R}}\,.
\end{equation}
Secondly, the main Gouy phase contribution comes from the distance from SRM to the telescope, because the Gouy phase changes faster near to the beam waist position, see Eq.~\ref{eq:psi}. For our analyses, a very short distance from the waist to the SRM of 10 m is assumed. Finally, a minimal distance $\sim$ 100\,m between the beam-expander mirrors helps to reduce the overall SRC length to $\sim$ \,250m whilst retaining a 6\,mm beam waist.

\begin{figure}[b]
    \centering
    \includegraphics[width=\columnwidth]{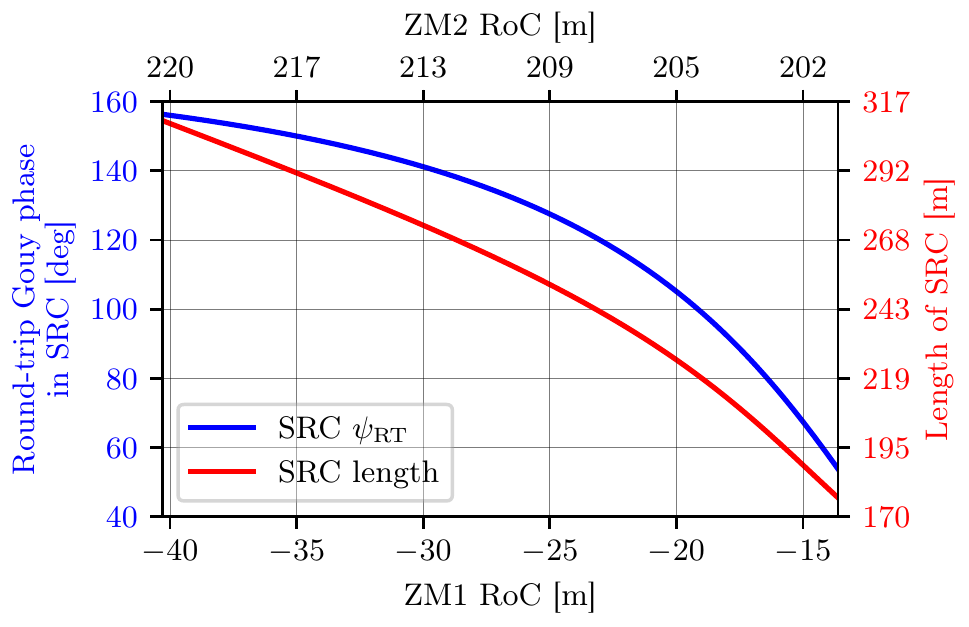}
    \caption{ZM1 and ZM2 RoC combinations yielding a 6 mm waist size where the position of this waist
    is less than 150 m from ZM1. The blue trace shows the round-trip Gouy phase, $\psi_{\mathrm{RT}}$, in the SRC. All the
    values in this trace satisfy the condition $20^{\circ} \leq \psi_{\mathrm{RT}} \leq 160^{\circ}$ so that every solution
    shown in this figure represents a stable SRC. The red trace gives the corresponding length of the SRC for each of the
    RoC combinations.}
    \label{fig:curved_z_solutions}
\end{figure}

The results shown in Figure~\ref{fig:curved_z_solutions} were obtained after a wide parameter space search. This figure shows that a stable SRC is possible, however, a relatively long SRC is required -- on the order of 250\,m on average - whilst the RoC of ZM1 is small relative
to the ZM2 RoC. We can trade off the achievable SRC length and the required curvature of ZM1, i.e. a shorter SRC can be achieved by reducing the radius of curvature of this mirror -- this ultimately comes down to a design choice, based on other design parameters outside the scope of this work.

\subsection{Curved telescope mirrors with a lens at the ITM}
\label{se:curved_zms_with_lens}
The previous solutions, shown in Figure~\ref{fig:curved_z_solutions}, lead to a relatively long SRC, $> 100$\,m.
In this section
we will investigate how adding a lens at the ITM can be used to reduce the length of the SRC and relax the requirements on the radius of curvature of ZM1.

We produced a set of animated plots of the round-trip Gouy phase to gain an intuition of how the solution regions evolve.
Similarly to \cref{se:curved_zms_no_lens}, the content shown here is based on these wide parameter searches and thus only shows the conclusions. We find short focal lengths ($f \lessapprox 100\,\mathrm{m}$) that result in a real waist beyond the ZM1 optic give solutions satisfying our requirements. A focal length of
$f = 75\,\mathrm{m}$ was chosen for a more in-depth analysis. Note that, any focal length comparable to this value will result in similar telescope behaviour but with slightly different solutions for the ZM1 and ZM2 curvatures and distances to the waist.

\begin{figure}[b]
    \centering
    \includegraphics[width=\columnwidth]{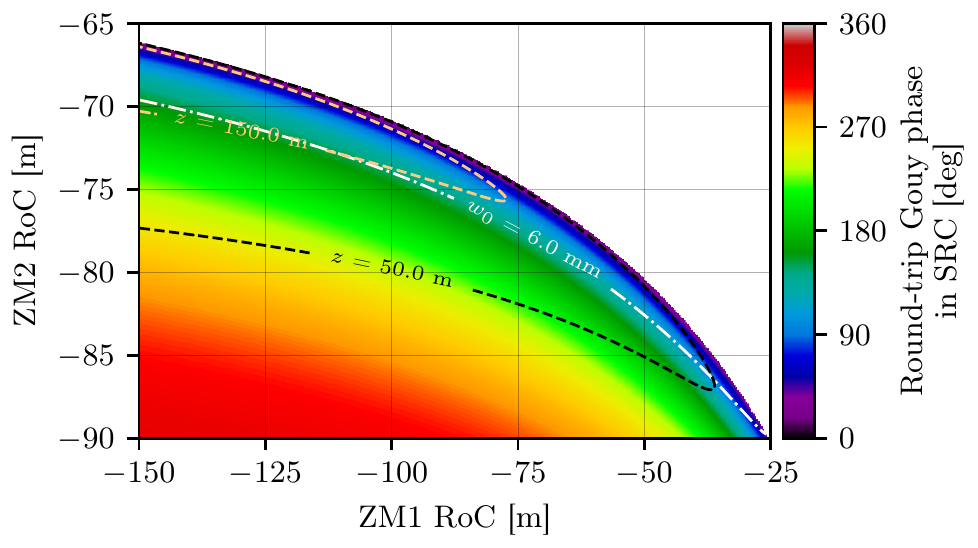}
    \caption{SRC round-trip Gouy phase for an ITM lens of focal length $f = 75\,\mathrm{m}$ and a distance between the Z
    mirrors of 50 m. Note that the solutions now require the curvatures of ZM1 and ZM2 to both be negative - this is because
    the ITM lens, of the focal length used here, is responsible for focusing the beam to a waist from the arm cavity. ZM1 collimates
    the beam going towards the beamsplitter, whilst ZM2 acts as a ``beam expander'' to prevent the beam (as propagated from the
    arm cavity) from focusing down to a waist too quickly.}
    \label{fig:curved_z_lens_rtgouy}
\end{figure}

\begin{figure}
    \centering
    \includegraphics[width=\columnwidth]{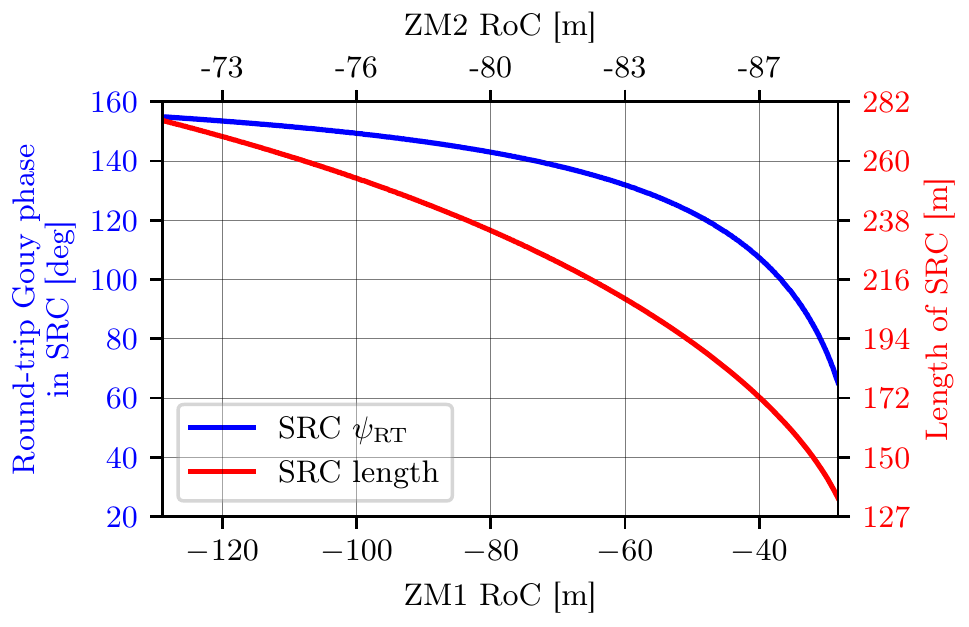}
    \caption{The key take-away from this plot is that by introducing a lens, shorter SRC lengths can be obtained along with the option of less demanding (i.e. larger) ZM1 RoCs. Taking $\rm{ZM1~RoC} = -40\,\mathrm{m}$ as an example data point, we see from this plot that this corresponds to an
    SRC length of approximately 170 m and a round-trip Gouy phase of approx. 110 degrees. Compare this to the same $\rm{ZM1~RoC}$ value on Figure
    \ref{fig:curved_z_solutions} which gives around 310 m and 155 degrees for these quantities, respectively.}
    \label{fig:curved_z_lens_solutions}
\end{figure}

Using this focal length, and a distance between the Z mirrors of 50\,m, Figure~\ref{fig:curved_z_lens_rtgouy} was produced -- giving the SRC round-trip Gouy phase over the ZM RoCs, with contours for the 6\,mm waist size and distances to the waist (from ZM1) of 50 and 150\,m overlaid on the plot. These contours, along with the color-map, frame the region which provides potential configurations for achieving a stable SRC of a suitable length. It is immediately apparent from this figure that the range of possible ZM1 curvatures which give solutions is much larger than for the configuration with no lens at the ITM. Here the lens takes the role of focusing the beam such that both ZM1 and ZM2 have negative RoCs and act together to collimate the beam from the arm cavity. By inspecting Figure~\ref{fig:curved_z_lens_rtgouy} we find that this solution region approximately corresponds to ZM1 RoC $ \in [-130\,\mathrm{m}, -30\,\mathrm{m}]$,
ZM2 RoC $ \in [-70\,\mathrm{m}, -90\,\mathrm{m}]$. This region is shown in Figure~\ref{fig:curved_z_lens_solutions}.

We can use Figure~\ref{fig:curved_z_lens_solutions} to pick a reference solution for the telescope parameters. One such solution set is given in Table~\ref{tab:tel_params_opt}, where the curvature combinations were chosen such that a relatively short SRC is obtained (important for ET-HF, see~\cref{se:constraints}) whilst the edges of the solution range are avoided (i.e. avoiding a near-unstable SRC). The resulting $g$-factor of the signal recycling cavity for this solution set is $g \sim 0.37$. In addition, this solution yields $> 99.9\%$ mode matching between the SRC and arm cavity. A discussion of the ET-HF results, also stated in Table \ref{tab:tel_params_opt}, is given in \cref{se:tel_designs_hf}.

\begin{table*}[tb]
\caption{Parameters of the telescopes chosen using Figure~\ref{fig:curved_z_lens_solutions}, with values for the beam size and accumulated Gouy phase associated with these given at each optic in the configuration. Note that the computed values have been given to 2 s.f. to avoid unnecessary precision at this stage. Where appropriate, values for LF and HF have been given separately. See Figure~\ref{fig:sol_trace} for a visual representation of these data for ET-LF. The focal length of the ITM lens, in both cases, is 75\,m.}
\begin{center}

\begin{tabular}{|c|c|l|c|c|c|c|ll}
\cline{1-7}
\textbf{Optic} &
  \multicolumn{2}{c|}{} &
  \textit{SRM} &
  \textit{BS} &
  \textit{ZM1} &
  \textit{ZM2} &
  \multirow{8}{*}{} &
  \multirow{8}{*}{} \\ \cline{1-7}
\multirow{2}{*}{RoC {[}m{]}} &
  \multicolumn{2}{c|}{LF} &
  -9410 &
  \multirow{2}{*}{inf} &
  \multirow{2}{*}{-50} &
  -82.5 &
   &
   \\ \cline{2-4} \cline{7-7}
 &
  \multicolumn{2}{c|}{HF} &
  -630 &
   &
   &
  -63.2 &
   &
   \\ \cline{1-7}
\multirow{2}{*}{Beam radius {[}mm{]}} &
  \multicolumn{2}{c|}{LF} &
  6.1 &
  6.2 &
  8.9 &
  30 &
   &
   \\ \cline{2-7}
 &
  \multicolumn{2}{c|}{HF} &
  6.3 &
  6.4 &
  8.3 &
  38 &
   &
   \\ \cline{1-7}
\textbf{Space} &
  \multicolumn{2}{c|}{} &
  \textit{SRM-BS} &
  \textit{BS-ZM1} &
  \textit{ZM1-ZM2} &
  \textit{ZM2-ITM} &
   &
   \\ \cline{1-7}
\multirow{2}{*}{Length {[}m{]}} &
  \multicolumn{2}{c|}{LF} &
  \multirow{2}{*}{10} &
  \multirow{2}{*}{70} &
  50 &
  \multirow{2}{*}{52.5} &
   &
   \\ \cline{2-3} \cline{6-6}
 &
  \multicolumn{2}{c|}{HF} &
   &
   &
  80 &
   &
   &
   \\ \hline
\multirow{2}{*}{Gouy phase {[}deg{]}} &
  \multicolumn{2}{c|}{LF} &
  7.5 &
  39 &
  5.3 &
  0.6 &
  \multicolumn{1}{c|}{\multirow{2}{*}{\begin{tabular}[c]{@{}c@{}}Total accumulated\\ Gouy phase {[}deg{]}\end{tabular}}} &
  \multicolumn{1}{l|}{52} \\ \cline{2-7} \cline{9-9}
 &
  \multicolumn{2}{c|}{HF} &
  4.8 &
  26 &
  4.9 &
  0.2 &
  \multicolumn{1}{c|}{} &
  \multicolumn{1}{l|}{36} \\ \hline
\end{tabular}

\end{center}
\label{tab:tel_params_opt}
\end{table*}

\begin{figure}
    \centering
    \includegraphics[width=\columnwidth]{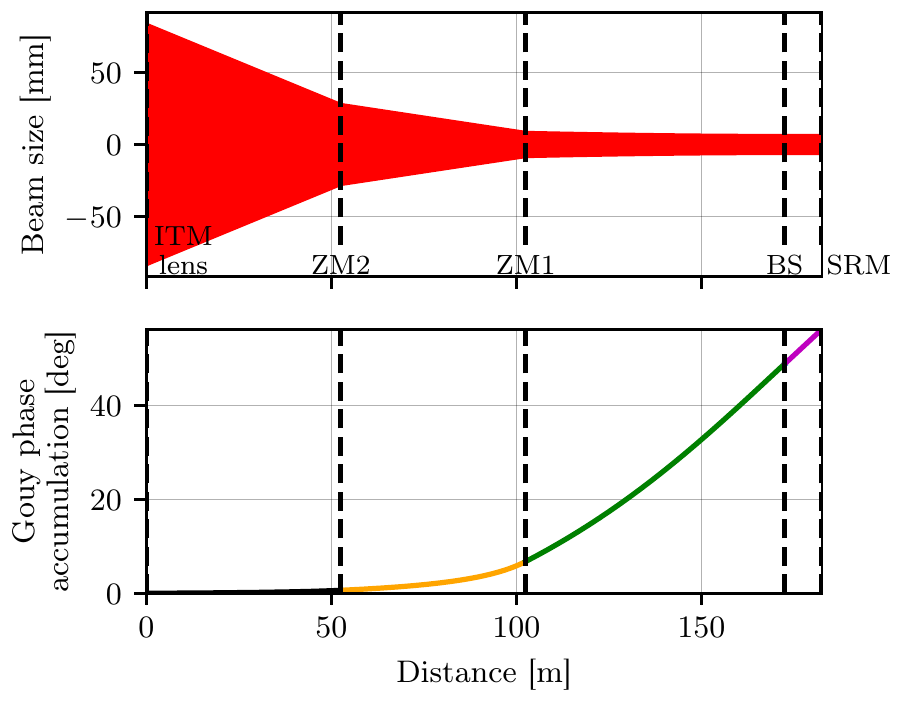}
    \caption{
    The ET-LF telescope design from the arm cavity to the SRM is shown; where the beam size and accumulated Gouy
    phase are plotted over the distance propagated through the SRC. Note that, whilst the beam sizes and telescope
    length differ slightly for ET-HF, the beam remains collimated in the ZM1 to SRM path; this is an important
    consideration for Figure~\ref{fig:zm1_bs_dist}.
    }
    \label{fig:sol_trace}
\end{figure}

\section{ET-HF arm telescope design}
\label{se:tel_designs_hf}

The ET-LF solution given in Table~\ref{tab:tel_params_opt} was used as a starting point for a similar analysis on the ET-HF arm telescope.
The results are then given in the same table, denoted with HF to distinguish the values from LF where appropriate. The larger beam size
(impinging on the arm cavity mirrors, see Table~\ref{tab:armcav_geometric_params}), and shorter wavelength of ET-HF, result in the requirement for
a longer telescope length when considering the same waist size target of 6\,mm. This increased length requirement can potentially be relaxed by decreasing this waist size target, however the required trade-off study is beyond the scope of this paper. The solution given for ET-HF in Table~\ref{tab:tel_params_opt} is optimised given this waist (and stable SRC) requirement. Note, also, that the ET-HF solution uses the same ITM lens focal length ($f = 75$\,m) as the ET-LF solution found in~\cref{se:curved_zms_with_lens}. The resulting $g$-factor of the signal recycling cavity for this solution set is $g \sim 0.65$. The solution stated also yields $> 99.9\%$ mode matching between the SRC and arm cavity.

\begin{figure}[b]
    \centering
    \includegraphics[width=\columnwidth]{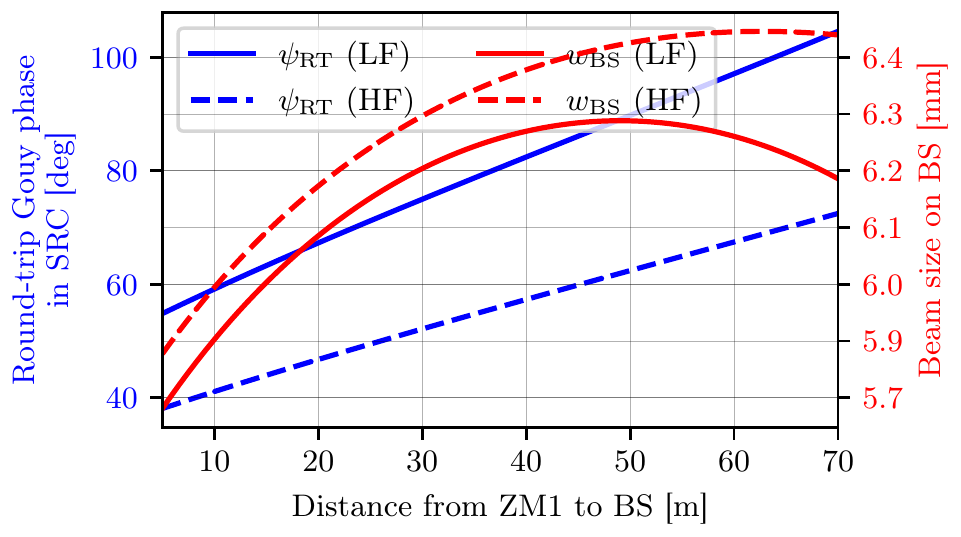}
    \caption{
    The accumulated round-trip Gouy phase in the SRC and radius of the beam impinging on BS for both ET-LF and ET-HF where
    the distance between the ZM1 and BS is decreased from its nominal value given in Table \ref{tab:tel_params_opt}. All
    other telescope parameters are kept constant.
    }
    \label{fig:zm1_bs_dist}
\end{figure}

Given that both the solutions for ET-LF and ET-HF result in a beam that is roughly collimated between ZM1 and the SRM (see Figure~\ref{fig:sol_trace}), we can alter the distance between BS and ZM1 without affecting the beam size on the beamsplitter by much more than a few hundred microns. This is demonstrated in Figure~\ref{fig:zm1_bs_dist}. Even by reducing this distance significantly, e.g. to 10\,m, a stable SRC can still be obtained for both ET-LF and ET-HF (with $\psi_{\mathrm{RT}} \approx 60$ and $\psi_{\mathrm{RT}} \approx 40$\,degrees, respectively); where the beam size on the beamsplitter would then be around 5.7\,mm and 5.9\,mm for ET-LF and ET-HF, respectively. Of particular importance to ET-HF, this could allow for a nominal reduction in the SRC length from 210\,m to around 150\,m whilst keeping the other telescope parameters, shown in Table~\ref{tab:tel_params_opt}, constant.

\section{Parameter sensitivity and mode matching}
\label{se:tolerancing}

\begin{figure*}
    \centering
    \subfloat[]{
        \includegraphics[width=\columnwidth]{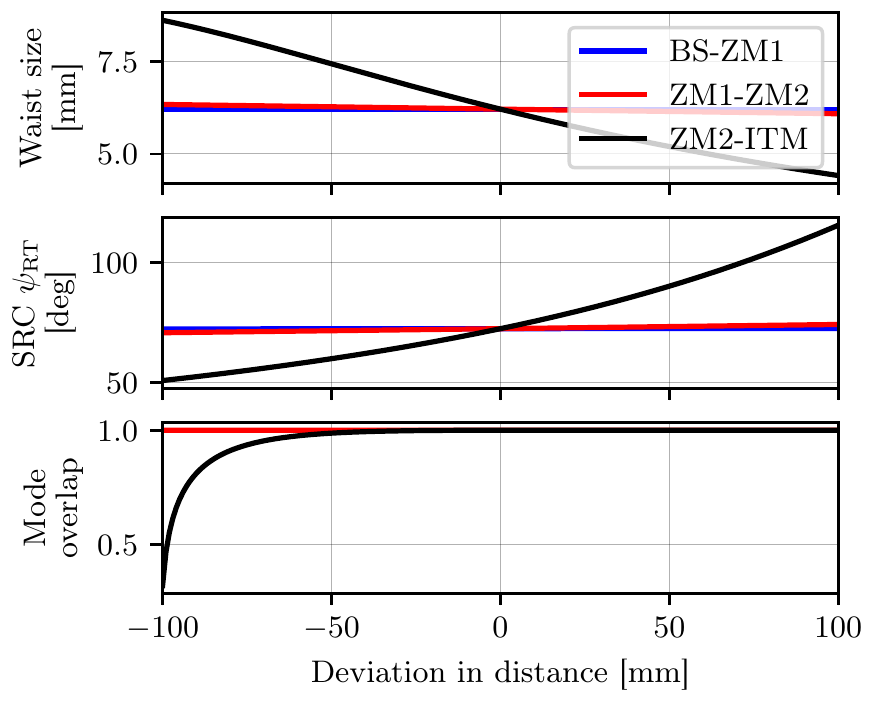}
        \label{fig:distance_devs}
    }
    \subfloat[]{
        \includegraphics[width=\columnwidth]{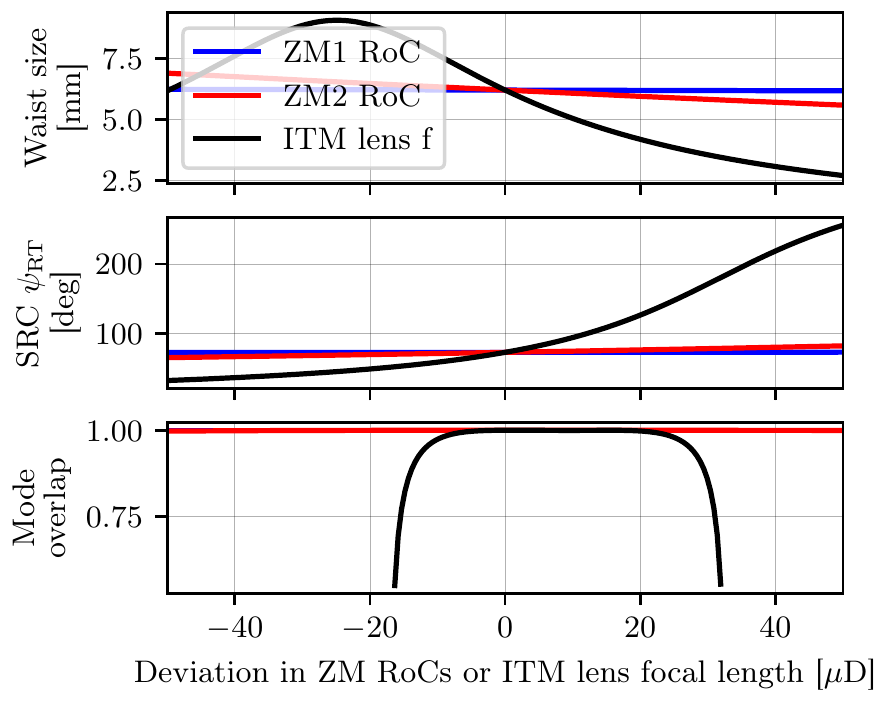}
        \label{fig:curv_devs}
    }
    \caption{SRC waist-size, round-trip Gouy phase and mode overlap with the arm cavity as functions of the key distances
    between optics, (a), and the curvatures of the telescope optics, (b). Each deviation is given in relative terms where
    a value of zero corresponds to the (ET-HF) baseline value given in Table~\ref{tab:tel_params_opt}. From (a) we can see
    that the distance between the ITM and ZM2 is the most critical length. Whilst, in (b), we find that the focal length
    of the ITM lens is the critical parameter in terms of the optic geometries. Note that the mode matching values in the lower
    sub-plot were computed via Equation~\eqref{eq:overlap}.}
    \label{fig:broad_devs}
\end{figure*}

Taking the results found in the end of~\cref{se:curved_zms_with_lens} as our baseline configuration, we can determine
the critical parameters of, for example, our ET-HF telescope design. These can be determined by deviating the key
free parameters of the system to observe the effect on the SRC waist size, round-trip Gouy phase and mode matching
to the arm cavity. Figure~\ref{fig:broad_devs} displays the results of such an analysis, where the left plots give the
aforementioned target parameters as a function of the distances between the optics whilst the right plots are based
on deviations in the radii of curvature of the Z mirrors and the focal length of the ITM lens.

The mode-matching quantity shown in Figure~\ref{fig:broad_devs} is defined by the ``overlap'' ($\mathcal{O}$) figure-of-merit~\cite{Bayer-Helms:84},

\begin{equation}
    \mathcal{O} = \frac{4|\Im{\{q_1\}}\,\Im{\{q_2\}}|}{|q_1^* - q_2|^2},
    \label{eq:overlap}
\end{equation}

\noindent \\
where $q_1$ and $q_2$ are the beam parameters being compared. In this case $q_1$ represents the eigenmode of the SRC
propagated to the arm cavity and $q_2$ is the arm cavity mode itself. This quantity returns values $\mathcal{O} \in [0, 1]$, where  unity indicates a full mode match between the two beam parameters and zero gives complete mode mismatch.

From Figure~\ref{fig:broad_devs} we can deduce, in terms of the optic geometries, that this telescope configuration is most sensitive to the focal length of the lens at the ITM. Thus the next step of this analysis will focus solely on the ITM lens focal length changes a result of thermal aberrations in the beam. Thermal lensing is investigated, in particular, due to it, potentially, being responsible for the largest effective
changes to the ITM lens focal length --- as shown at the end of the next section.

\subsection{Mode matching in the presence of thermal lensing}
\label{se:thermal_lens}

Surface deformation and refractive index differentials, caused by temperature distributions in the mirror substrates, lead to thermal aberrations (lensing) in the beam in ground-based GW detectors~\cite{refId0}. This thermal lensing
results in mode mismatches between the arm and recycling cavities. In terms of the optics present in our configuration, the thermal lensing acts to modify the effective focal length of the ITM lens, thereby altering the geometry of the beam in the signal recycling cavity (see the solid traces in Figure~\ref{fig:broad_devs} for this effect in a broad sense). To minimise these distortions, adaptive optics are required~\cite{Brooks:s, Rocchi_2012}. This is one of the tasks which could be performed by the arm telescopes of the ET detectors, thus in this section we will quantify the required deviations to the telescope mirror curvatures, for recovering mode matching to the arm cavities, in the presence of varied thermal lens focal lengths. In this case the ET-HF telescope (plus arm cavity) configuration is used, as this detector is designed to operate at high power~\cite{ET2020, ET-D} where thermal lensing will be more prevalent.

\begin{figure}
    \centering
    \includegraphics[width=\columnwidth]{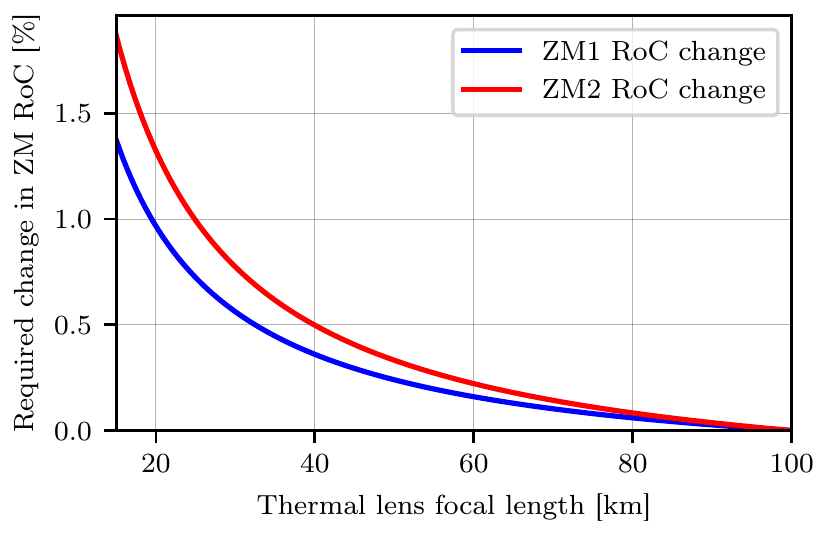}
    \caption{
    Simultaneous changes in Z mirror RoCs required for recovering ``complete'' mode matching from the SRC to the arm cavity. At the extreme of $f_\mathrm{th} = 15\,\mathrm{km}$ on this plot, the required deviation in the RoC of ZM1 is approximately 1.4\% whilst for ZM2 it is 1.9\%.
    }
    \label{fig:th_lens_mm}
\end{figure}

Figure~\ref{fig:th_lens_mm} quantifies the necessary modifications to the radii of curvature of ZM1 and ZM2 in order to recover $> 99.9\%$ mode matching of the SRC to the arm cavity; for an assumed range of thermal lens focal lengths of $f_{\mathrm{th}} \in [100\,\mathrm{km}, 15\,\mathrm{km}]$.
Note that for a strong focal length of 15\,km from the thermal lens, the effective focal length of the ITM  reduces to $f \approx 74.63\,\mathrm{m}$; i.e. a deviation of about 0.5\% from the target value of 75 m noted in~\cref{se:tolerancing}. This focal length distortion results in a mode mismatch, between the arm cavity and signal recycling cavity, of approximately 20\%. At this extreme thermal lens, the modifications to the Z mirror radii of curvature indicated by Figure~\ref{fig:th_lens_mm} reduce this mode mismatch to effectively $0\%$.

\section{Conclusion}
\label{se:conclusion}

In this paper we investigated arm telescope configurations, for both ET-LF and ET-HF, which are suitable for the optical layout
in the ET 2020 design update \cite{ET2020}. These telescope configurations are motivated by smaller beams on the optics in the central part of the interferometer.
The beam expanders also provide the ability to steer the
ET-HF beam around the ET-LF ITM suspension systems; with the added benefits of decoupling the angle of incidence on the beamsplitter from the beam axes in the
interferometer arms. Our requirements for this telescope can be summarised
as targeting a 6\,mm waist size positioned at the main beamsplitter whilst maintaining a stable SRC (quantified approximately as having
a length of the same order as the Rayleigh range of the beam). Further details on these requirements were given in~\cref{se:constraints}.

We demonstrated that it is possible to achieve a stable SRC, of a sensible length, with telescopes in the arms of both
the ET-LF and ET-HF interferometers. Reducing the length of SRC, in accordance with~\cite{PhysRevD.101.082002}, can be
attained via the introduction of a lens at the ITM for pre-focusing the beam from the arm cavity. Our baseline solutions
for such a configuration are given in Table~\ref{tab:tel_params_opt}. Further reductions to
the length of the SRC, whilst changing the spot size on the beamsplitter by only a few hundred microns, are possible via
decreasing the distance from ZM1 to BS - see Figure~\ref{fig:zm1_bs_dist} for details.

Our baseline configuration for ET-HF was analysed in~\cref{se:tolerancing} where we found that the focal length of the ITM lens is the critical
parameter in terms of the sensitivity for mode matching and SRC stability. However, in~\cref{se:thermal_lens}, we found that
effective changes in this focal length due to thermal lensing can be compensated with actuation on the curvatures of the
telescope mirrors. In particular, we saw that the mode mismatch (of approximately 20\%) due to a strong thermal lens, with $f_{\mathrm{th}} \sim 15\,\mathrm{km}$, can be fully corrected with changes of approximately 1.4\% and 1.9\% in the RoCs of ZM1 and ZM2, respectively.

The results presented here provide evidence for the feasibility of beam-expander telescopes in the interferometer arms of ET and provide essential input for trade-off studies of the optical layout. Further studies are required to study other aspects of this setup, in particular the effects of astigmatism in specific telescope implementations, and the possible negative impact on the contrast defect due to having the telescopes in a configuration that allows differential beam tuning.

\bibliography{bibliography}

\end{document}